\documentclass{PoS}

\newcommand{\msub}[1]{\ensuremath _{\mbox{\tiny #1}}}

\title{High temperature quark localization by Polyakov loops}

\ShortTitle{High temperature quark localization by Polyakov loops}

\author{\speaker{Tam\'as G.\ Kov\'acs}\thanks{Supported by EU Grant
    (FP7/2007-2013)/ERC N$^o$208740. We thank the Budapest-Wuppertal
    collaboration for providing us their code for generating the dynamical
    configurations used in this work.}  \mbox{\hspace{0.2ex} and} Ferenc
  Pittler$^\dagger$ \\ 
    Department of Physics, University of P\'ecs \\ 
    H-7624 P\'ecs, Ifj\'us\'ag \'utja 6, Hungary}

 \author{Falk Bruckmann\thanks{Supported by DFG Grant
     (BR-2872/4-2).} \hspace{0.5ex} and Sebastian Schierenberg$^\ddagger$
   \\ Institut f\"ur Theoretische Physik \\ D-93040 Regensburg, Germany}

\abstract{We study the low eigenmodes of the overlap and staggered Dirac
  operator at high temperature. We show that the recently found localized
  quark modes obeying Poisson statistics are connected to physical gauge field
  objects with their size and density scaling in the continuum limit. The
  localized modes are also strongly correlated with large fluctuations of the
  Polyakov loop. Based on that we construct a random matrix model of the low
  Dirac modes inspired by dimensional reduction. Our model reproduces the
  Poisson to random matrix transition seen in the lattice Dirac spectrum.}

\FullConference{ The XXIX International Symposium on Lattice Field Theory -
  Lattice 2011\\ July 10-16, 2011\\ Squaw Valley, Lake Tahoe, California}

\begin{document}

\section{Introduction}

It is by now well established that the high temperature QCD Dirac spectrum has
a remarkable feature, a transition from localized to delocalized modes
\cite{GarciaGarcia:2006gr}-\cite{Kovacs:2010wx}. It seems to be a generic
feature of non-Abelian gauge theories in four dimensions that at high
temperature, where chiral symmetry is restored, the lowest part of the Dirac
spectrum consists of localized modes exhibiting Poisson statistics. Higher up
in the spectrum there is a cross-over to delocalized modes and random matrix
statistics persisting throughout the bulk of the spectrum. This behavior,
reminiscent of Anderson localization, was first seen in the quenched $SU(2)$
theory with the overlap Dirac operator \cite{Kovacs:2009zj} and later with the
staggered Dirac operator \cite{Kovacs:2010wx}.

The detailed mechanism behind quark localization is still not known. In
particular, it is not clear whether there are any easily identifiable gauge
field configurations that are capable of binding and localizing the lowest
quark eigenmodes. In the present paper, which is mostly based on
Ref.\ \cite{Bruckmann:2011cc}, we present some new findings concerning the
nature of these gauge field objects. The lattice data we use in support of
these results come from various simulations including overlap and staggered
Dirac spectra in quenched $SU(2)$ gauge backgrounds and stout smeared
staggered spectra in $SU(3)$ backgrounds with 2+1 flavors of dynamical
quarks. The results of the dynamical simulations have not been published yet,
some preliminary results appear in the presentation by F.\ Pittler at this
Conference.

\section{Correlation between overlap and staggered modes}

\begin{figure}
\begin{center}
\includegraphics[width=0.75\columnwidth,keepaspectratio]{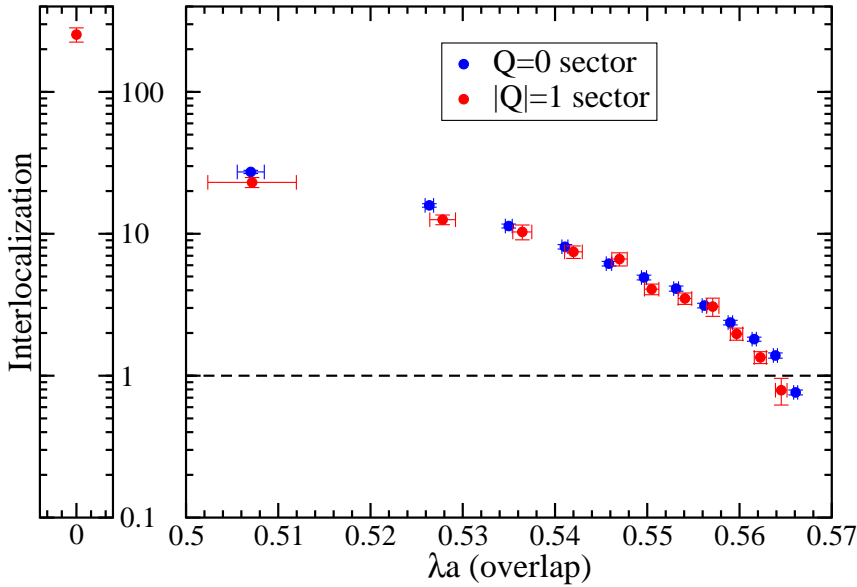}
\caption{\label{fig:interloc} The average interlocalization (see Eq.\ (2.1))
  of the lowest overlap and staggered eigenmodes as a function of the overlap
  eigenvalue in the topological charge sectors $Q=0$ and $|Q|=1$ }
\end{center}
\end{figure}

If the low Dirac modes are localized on specific gauge objects then in a given
gauge field background one expects the localized modes to occur around similar
locations irrespective of the particular discretization of the Dirac operator.
To test this we compared the lowest twelve eigenmodes of the staggered and the
overlap Dirac operator \cite{Neuberger:1997fp} on the same set of gauge field
backgrounds. For quantifying the spatial overlap between two eigenmodes we
defined the quantity
\begin{equation}
I = V \; \sum_x \; |\phi\msub{stag}(x)|^2 \; |\psi\msub{ov}(x)|^2
   \label{eq:interl}
\end{equation}
that we call {\em interlocalization}. For non-overlapping modes $I$ can be
close to zero while for exactly identical modes $I$ coincides with the inverse
of the participation ratio. Here the eigenmodes are always assumed to be
normalized. To have a meaningful comparison we first paired the lowest overlap
eigenmode with the staggered mode that had the largest interlocalization with
it. Then the second lowest overlap mode was paired with the remaining
staggered mode with maximal interlocalization with it and so on. After this
pairing was done configuration by configuration, we computed the average
interlocalization for the lowest twelve overlap modes and the corresponding
staggered modes. In Figure \ref{fig:interloc} we plot this quantity as a
function of the overlap eigenvalue both in the $Q=0$ and the $|Q|=1$
topological sectors.

Indeed, in both topological sectors the lowest staggered and overlap
eigenmodes have substantial overlaps that diminishes higher up in the
spectrum. We note that putting Gaussian random amplitudes for the two vectors
yields an interlocalization of unity. We can thus conclude that there are some
objects in the gauge configurations that bind localized eigenmodes of both the
overlap and the staggered Dirac operators. We note here that similarities
between overlap and staggered spectra had already been noted for the Schwinger
model \cite{Durr:2003xs} and for QCD \cite{Durr:2004as}, but in our case the
similarity also extends to the spatial structure of the eigenmodes.

\section{Scaling in the continuum limit}

\begin{figure}
\begin{center}
\includegraphics[width=0.75\columnwidth,keepaspectratio]{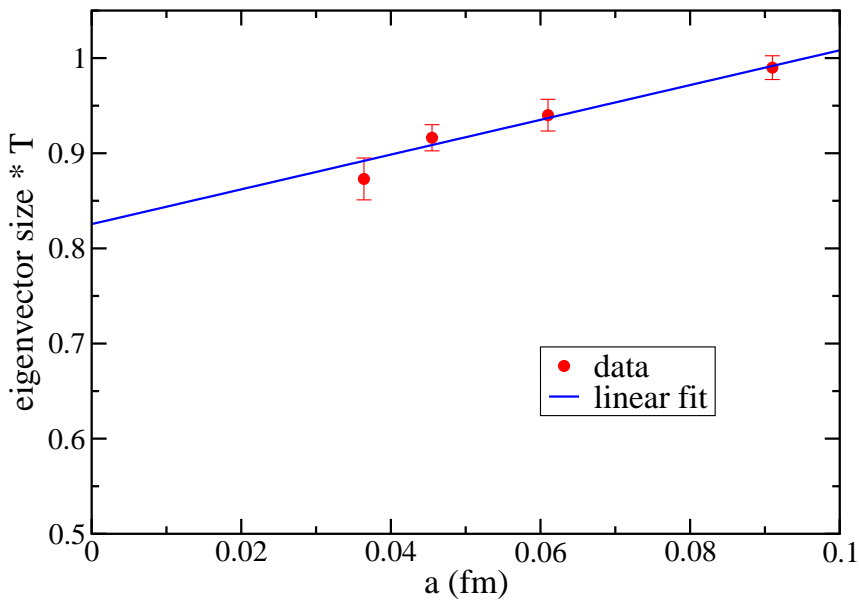}
\caption{\label{fig:evsize} The average linear size of the localized low
  eigenmodes of the overlap Dirac operator as a function of the lattice
  spacing. The eigenmode size is measured in units if inverse temperature. 
}
\end{center}
\end{figure}

Since the lowest eigenmodes are localized to within a few lattice spacings one
might be inclined to identify these gauge field objects with some sort of
``dislocations'', unusually large fluctuations in the gauge fields on the
scale of the lattice spacing. To explore this possibility we repeated the
simulations on three finer grids and, based on the participation ratio,
computed the linear size of the localized eigenmodes \cite{Kovacs:2010tu}. All
these simulations were done at the same physical temperature $(T=2.6T_c)$ and
in spatial boxes of the same physical size with the lattice spacing set by the
critical temperature. The results are shown in Figure \ref{fig:evsize} where
we plot the eigenvector size in units of the inverse temperature. A priori it
is not clear how to extrapolate this quantity to the continuum limit, but to
guide the eye we included a linear extrapolation. It is obvious that using any
sensible extrapolation to the continuum limit yields a non-zero value, in fact
a number of order unity. This clearly rules out dislocations as candidates for
binding the localized modes. Moreover, it seems that the spatial size of the
localized modes is set by the box size in the temporal direction.

To obtain further information concerning the nature of the gauge objects
binding the localized modes we can also look at their density and in
particular how that scales in the continuum limit.  Since there is no sharp
boundary between the localized and the delocalized modes in the spectrum, it
is not straightforward to count the number of localized modes and to define
their density. For instance in terms of the level spacing distribution,
$P(s)$, the exponential distribution, $P(s)=\exp(-s)$, characterizing localized
modes continuously changes into the Wigner surmise as we go up in the spectrum
and reach the regime of delocalized modes. If this path of deformation is
universal, which seems to be the case here, one can arbitrarily choose a
``standard'' distribution somewhere between the exponential and the Wigner
surmise. One can then call those eigenmodes localized that are below the point
in the spectrum where the level spacing distribution reaches this standard
``in-between'' distribution.

\begin{figure}
\begin{center}
\includegraphics[width=0.75\columnwidth,keepaspectratio]{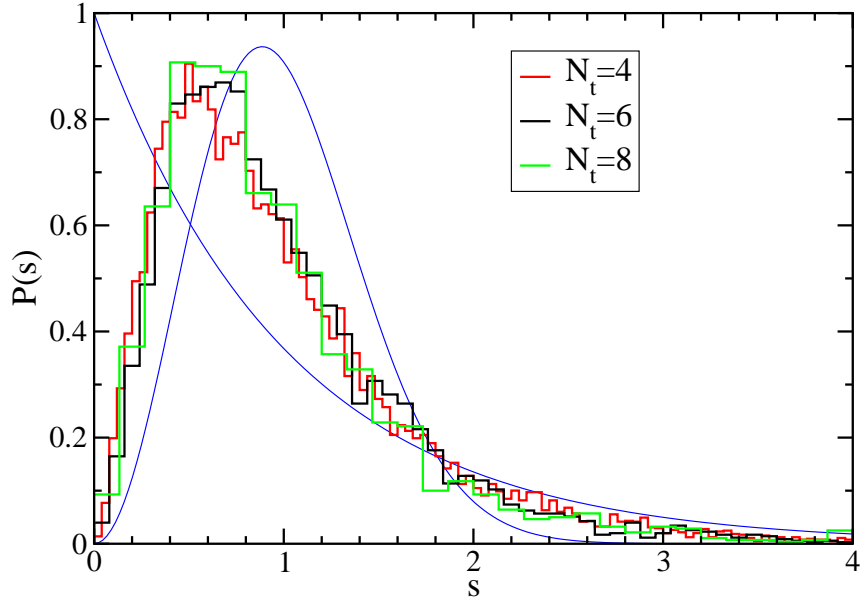}
\caption{\label{fig:lsd_scaling} The unfolded level spacing distribution
  computed from the eigenvalues between the 10$^{th}$ and 20$^{th}$. The
  histograms correspond to three ensembles with the same physical parameters,
  differing only in the lattice spacing. The two curved lines indicate the
  exponential and unitary Wigner surmise distributions corresponding to the
  lowest part and the bulk of the spectrum.  }
\end{center}
\end{figure}

In Figure \ref{fig:lsd_scaling} we show the unfolded level spacing
distribution computed from eigenvalues between number 10 and 20 of the
staggered Dirac operator in gauge field backgrounds generated with 2+1 flavors
of dynamical fermions with physical masses using the action in
Ref.\ \cite{Aoki:2006br}. The three curves correspond to three different
ensembles with $N_t=4,6$ and 8, at the same physical temperature and with the
same physical spatial box size. The two curved lines indicate the exponential
and the unitary Wigner surmise distributions observed in the lowest part and
the bulk of the spectrum. The histograms are exactly on top of one another
illustrating that the deformation of the distribution occurs along the same
universal path regardless of the lattice spacing. Since the matching
histograms correspond to the same slice of the spectrum (between eigenvalues
10 and 20) in all three cases, we can conclude that according to our
definition all these ensembles have the same number of localized modes per
configuration. This in turn implies that the physical density of localized
modes is also the same since the physical three-volumes of the ensembles were
chosen to be identical.

We already know that localized low Dirac modes are bound to some gauge objects
that have a fixed physical size and physical density in the continuum
limit. It would be tempting to identify these as finite temperature
topological objects, calorons with their constituent monopoles. Based on the
zero modes of the overlap Dirac operator on these gauge backgrounds we can
estimate that at this temperature there are on average about 0.01 topological
objects (calorons or anti-calorons) per cubic fermi. This estimate also
depends on the assumption that since the gas of topological objects is dilute,
they are uncorrelated. On the other hand, the density of localized modes turns
out to be about 1 per cubic fermi. The two orders of magnitude discrepancy
between the density of calorons and localized modes makes it impossible
that a dilute gas of topological objects can explain the localized Dirac modes.

\section{Connection to Polyakov loops}

To gain further intuition concerning the origin of these low Dirac modes we
note that the ``thinning out'' of Dirac modes around zero at high temperature,
needed for chiral restoration, can be qualitatively understood by considering
the lowest Matsubara frequency. Indeed, due to the anti-periodic temporal
boundary condition the lowest free quark modes are shifted away from zero by
an amount $\propto T$. If the gauge field is ``turned on'' the effective
temporal boundary condition for the quarks is the combination of the
anti-periodic boundary condition and the Polyakov loop. At high temperature
the latter has a positive real expectation value, however, it can locally
fluctuate. Fluctuations of the Polyakov loop can in principle locally lower
the effective Matsubara frequency and shift some Dirac eigenvalues towards
zero. If this is the case one expects the lowest eigenmodes to be localized at
places where the Polyakov loop has large fluctuations.

\begin{figure}
\begin{center}
\includegraphics[width=0.75\columnwidth,keepaspectratio]{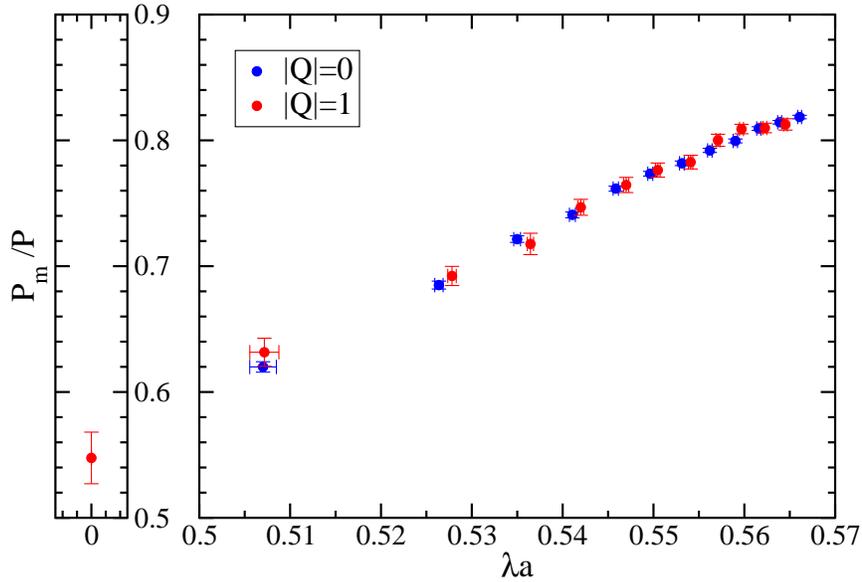}
\caption{\label{fig:poly_overlap} The weighted average of the Polyakov loop
  versus the overlap eigenmode the magnitude of which was used for the
  weighting. The weighted average is normalized by the simple average.}
\end{center}
\end{figure}

Whether that is the case can be tested by comparing the weighted average of
the Polyakov loop to the simple average. We choose the weight on each site to
be the magnitude of a normalized Dirac wave function on that site. In this way
to each Dirac eigenmode we can associate a weighted Polyakov loop with the
formula
\begin{equation}
 P_m = \sum_x P(x) |\psi_m(x)|^2.
\end{equation}

In Figure \ref{fig:poly_overlap} we plot this quantity normalized by the
simple average Polyakov loop for the pure $SU(2)$ gauge theory for the first
few overlap eigenvalues. It is clear that for the low Dirac modes the weighted
average is significantly lower than the simple average. This demonstrates that
the localized Dirac modes are indeed peaked at locations where the Polyakov
loop has large local fluctuations towards smaller and even negative values
that significantly lower the effective Matsubara frequency.

To test the validity of this picture in a more quantitative fashion we also
explored an effective random matrix model inspired by this picture. Splitting
the Dirac operator into a temporal and a spatial part we choose a basis that
diagonalizes the temporal part with diagonal matrix elements 
\begin{equation}
 \lambda(x) a = \sin\frac{\pi-\phi(x)}{N_t},
\end{equation}
where $\lambda(x)$ is the effective local lattice Matsubara frequency and
$\phi(x)$ corresponds to the local phase of the Polyakov loop at $x$. The
extra phase $\pi$ in this formula represents the anti-periodic temporal
boundary condition. We further assume that this three-dimensional array of
local Matsubara modes interact through nearest neighbor interactions in all
spatial directions. These random spatial interactions are meant to capture the
effect of spatial gauge couplings in the Dirac operator. This sparse random
matrix model has a dimensionally reduced three-dimensional structure and is
similar to the Anderson model based on the tight binding approximation.  The
random on-site terms are the local Matsubara frequencies and the nearest
neighbor hopping terms are the effective interaction terms between neighboring
Matsubara modes. This model has only a few parameters to be fixed. We
experimented with some parameter sets inspired by actual lattice data and
found that for a range of parameters this random matrix ensemble reproduces
the qualitative features of the lattice Dirac operator. The lowest part of the
spectrum consists of localized modes with exponentially distributed nearest
neighbor level spacings and the bulk of the spectrum has delocalized modes
with random matrix statistics. For details we refer the reader to
Ref.\ \cite{Bruckmann:2011cc}.

\section{Conclusions}

In the present paper we studied how the localization of the lowest quark modes
occurs in QCD at high temperature. We showed that the location of the
eigenmodes is robust with respect to different discretizations of the Dirac
operator. We demonstrated this by comparing staggered and overlap eigenmodes
on the same gauge backgrounds. Using simulations with four different lattice
spacings we found that the physical size of the localized modes has a non-zero
continuum limit which is of the order of the inverse temperature. 
This implies that localized modes are not bound to dislocations, rather they
are connected to gauge field objects the extension of which is controlled by the
temporal box size. We further demonstrated that the physical density of these
objects does not depend on the lattice spacing. Their density turned out to be
about two orders of magnitude larger than the density of topological objects.
We found strong correlations between the locations of low Dirac
modes and those of large fluctuations of the Polyakov loop. Based on that we
proposed a dimensionally reduced sparse random matrix model of
localization. It would be interesting to explicitly identify the gauge field
objects responsible for localization. Monopoles and dyons are good candidates
for that, but some non-trivial correlations between different objects would be
needed \cite{Ilgenfritz:1994nt},\cite{Unsal:2007jx}.


\begin{thebibliography}{99}

\bibitem{GarciaGarcia:2006gr}
  A.~M.~Garcia-Garcia and J.~C.~Osborn,
  Phys.\ Rev.\  D {\bf 75}, 034503 (2007)
  [arXiv:hep-lat/0611019];
  A.~M.~Garcia-Garcia, J.~C.~Osborn,
  Nucl.\ Phys.\  {\bf A770}, 141-161 (2006).
  [hep-lat/0512025].



\bibitem{Kovacs:2009zj}
  T.~G.~Kovacs,
  Phys.\ Rev.\ Lett.\  {\bf 104}, 031601 (2010)
  [arXiv:0906.5373 [hep-lat]].

\bibitem{Kovacs:2010wx}
  T.~G.~Kovacs, F.~Pittler,
  Phys.\ Rev.\ Lett.\  {\bf 105}, 192001 (2010).
  [arXiv:1006.1205 [hep-lat]].

\bibitem{Bruckmann:2011cc}
  F.~Bruckmann, T.~G.~Kovacs, S.~Schierenberg,
  Phys.\ Rev.\  {\bf D84}, 034505 (2011).
  [arXiv:1105.5336 [hep-lat]].


\bibitem{Kovacs:2010tu}
  T.~G.~Kovacs, F.~Pittler,
  PoS {\bf LATTICE2010}, 195 (2010).
  [arXiv:1011.3175 [hep-lat]].



\bibitem{Neuberger:1997fp}
  H.~Neuberger,
  Phys.\ Lett.\  {\bf B417}, 141 (1998), hep-lat/9707022;
%
  H.~Neuberger,
  Phys.\ Lett.\  {\bf B427}, 353 (1998), hep-lat/9801031].

\bibitem{Durr:2003xs}
  S.~D\"urr, C.~Hoelbling,
  Phys.\ Rev.\  {\bf D69}, 034503 (2004),
  hep-lat/0311002.

\bibitem{Durr:2004as}
  S.~D\"urr, C.~Hoelbling, U.~Wenger,
  Phys.\ Rev.\  {\bf D70}, 094502 (2004),
  hep-lat/0406027.



\bibitem{Aoki:2006br}
  Y.~Aoki, Z.~Fodor, S.~D.~Katz, K.~K.~Szabo,
  Phys.\ Lett.\  {\bf B643}, 46-54 (2006).
  [hep-lat/0609068].

\bibitem{Ilgenfritz:1994nt}
  E.~-M.~Ilgenfritz, E.~V.~Shuryak,
  Phys.\ Lett.\  {\bf B325}, 263-266 (1994).
  [hep-ph/9401285].

\bibitem{Unsal:2007jx}
  M.~Unsal,
  Phys.\ Rev.\  D {\bf 80}, 065001 (2009)
  [arXiv:0709.3269 [hep-th]].





\end{thebibliography}
\end{document}